\documentclass[aps,prd,preprint,nofootinbib,superscriptaddress]{revtex4-1}

\usepackage{color}
\usepackage{amssymb}  
\usepackage{amsthm}
\usepackage{amsmath}
\usepackage{graphicx}
\usepackage{soul}

\newcommand{\bel}[1] {\begin{equation}\label{#1}}
\newcommand{\beal}[1] {\begin{eqnarray}\label{#1}}
\newcommand{\expec}[1] {\langle {#1} \rangle}

\def\gra{\eta} 
\def\hs{\zeta} 
\def\poten{V}
\def\pot{ U}
\def\fn{\phi}
\def\Tr{\rm Tr}
\def\({\left(}
\def\){\right)}
\def\[{\left[}
\def\]{\right]}
\def\det{{\rm det}}
\def\nn{\nonumber}
\def\ee{\end{equation}}
\def\eea{\end{eqnarray}}
\def\nn{\nonumber \\}

\def\field{\phi}
\def\be{\begin{equation}}
\def\bea{\begin{eqnarray}}

\begin{document}


\title{Counting Vacua in Random Landscapes}

\author{Richard Easther} 
\email[]{r.easther@auckland.ac.nz}
\affiliation{Department of Physics, University of Auckland, Private Bag 92019, Auckland, New Zealand}
\author{Alan H. Guth}
\email[]{guth@ctp.mit.edu}
\affiliation{Center for Theoretical Physics \& Department of Physics, Massachusetts Institute of Technology, Cambridge, MA 02139, USA}
\author{Ali Masoumi}
\email[]{ali@cosmos.phy.tufts.edu}
\affiliation{Institute of Cosmology, Department of Physics and Astronomy,
Tufts University, Medford, MA 02155, USA}


\date{\today}

\begin{abstract} 
\noindent It is speculated that the correct theory of fundamental physics includes a large {\em landscape\/} of states, which can be described as a potential which is a function of $N$ scalar fields and some number of discrete variables. The properties of such a landscape are crucial in determining  key cosmological parameters including the dark energy density, the stability of the vacuum, the naturalness of  inflation and the properties of the resulting perturbations, and the likelihood of bubble nucleation events.  We codify an approach to landscape cosmology based on  specifications of the overall form of the landscape potential and illustrate this approach with a detailed analysis of the  properties of $N$-dimensional Gaussian random landscapes. We clarify the correlations between the different matrix elements of the Hessian at the stationary points of the potential. We show that  these potentials generically contain a large number of minima. More generally, these results elucidate how random function theory is of central importance  to this approach to landscape cosmology, yielding results that differ substantially from those obtained by treating the matrix elements of the Hessian as independent random variables. 
\end{abstract}


 \maketitle

\section{Introduction}

For over a decade,  considerations motivated by  flux-compactified string vacua \cite{Bousso:2000xa,Feng:2000if,Banks:2003sx,Polchinski:2006gy,ArkaniHamed:2006dz}  have suggested that fundamental physics may be specified within a {\em landscape\/}, a highly complex, multidimensional scalar potential.  From this perspective the search for a unique theory of everything yields an apparent theory of anything: a vast range of possible configurations of ``low energy'' physics. An immediate corollary of this development is that if such a landscape emerges from  fundamental physics, then anthropic reasoning may be central to understanding the observed properties of our universe. 

Unfortunately, the string landscape itself appears to be so complex that quantitative explorations of its properties are computationally intractable. However,  an alternative perspective is to treat the landscape as a realization of an  $N$-dimensional random function $\poten(\bar \field)$ drawn from a specified distribution.\footnote{This does not imply that the form of the landscape is  arbitrary, but rather that it consists of a sufficient number of largely uncorrelated terms that it can be treated probabilistically and in this scenario the most natural choice of distribution will be Gaussian, motivated by the central limit theorem.}  This distribution is fixed by the hypothesized {\em architecture\/} of the landscape. The critical observation  underpinning this approach is that in many scenarios the tools of random function / random matrix theory will yield the distributions of key cosmological observables within this landscape. Physically, this approach is reasonable if we have grounds to believe that a landscape potential is a  superposition of many largely independent terms, rather than (for instance) an almost periodic function with strong long-range correlations. 

An early step in this direction was taken in Ref.~\cite{Aazami:2005jf}, which posited that the ensemble of Hessian matrices associated with the stationary points in a generic landscape could be described by a set of symmetric matrices with elements  chosen from independent identical Gaussian distributions. This ensemble is called the Gaussian Orthogonal Ensemble (GOE).  From this perspective it appears that relative to saddle points,  minima are super-exponentially rare, as they correspond to large fluctuations from the Wigner semi-circle eigenvalue distribution \cite{Dean:2006wk}. Similar arguments were made about the critical points in a general four-dimensional ${\cal{N}} = 1$ supergravity  \cite{Marsh:2011aa}. 

Conversely,  Battefeld {\em et al.} \cite{Battefeld:2012qx}  gave a semi-analytic treatment of the properties of Hessian matrices associated with minima of a ``softly bounded'' landscape, looking at Hessians derived directly from explicitly constructed random functions, in this case  finite sums of Fourier terms. Stability and distribution of the vacuum energies of this theory was studied in \cite{Masoumi:2016eqo}. These potentials are naturally bounded below and while minima are outnumbered by saddles, they are not super-exponentially rare.  Treating the potential as a random function also facilitates analyses of the vacuum stability, and the likelihood of tunneling has been studied in  polynomial \cite{Greene:2013ida, Aravind:2014aza,Dine:2015ioa,Dine:2015szg} and  bounded Fourier landscapes  \cite{Masoumi:2016eqo}. 

More generally, in the large-$N$ limit,  distributions of parameters associated with random functions and random matrices are often well-defined, potentially  transforming the complexity of the landscape into a predictive tool. For simpler models with many non-interacting fields this approach has led to predictions for the mass-spectrum of N-flation \cite{Easther:2005zr} and the perturbation spectra of many-field scenarios \cite{Easther:2013rva,Price:2014ufa}.  However, the key insight of this paper is to elucidate how a full understanding of the properties of critical points in a generic landscape with nontrivial couplings between the  fields will require random {\em function\/} theory and not just random matrix theory, which implicitly treats the ensemble of Hessian matrices at extrema as uncorrelated matrices with uncorrelated elements.

 The primary goal of this paper is to fully develop random function theory as a tool for understanding  landscape models, extending the methods of Ref.~ \cite{Bardeen:1985tr} to $N$ dimensions and understanding the approach to the large-$N$ limit \cite{Bray:2007tf}.  In doing so we  categorize correlations between elements of the Hessian matrices, and elucidate the ways in which properties of extrema are correlated with the value of the potential. These correlations can be partly understood on purely topological grounds and highlight the information  discarded by analyses which treat elements of the Hessian as independent and identically distributed variables. 

As an illustrative example, perhaps the two simplest possible architectures are i) a set of uncoupled, self-interacting fields and ii) a potential $\poten(\bar\field)$ which is an $N$-dimensional isotropic, Gaussian random field, where $\bar\field$ denotes the $N$ independent scalar fields, and $\poten$ is   a map from $\mathbb{R}^N$ to $\mathbb{R}$.  The distribution of possible values of the cosmological constant, $p(\Lambda)$, in a specific realization of the landscape is synonymous with the  distribution of values of  $\poten(\bar\field)$ at  minima of the potential. The expected value of $\Lambda$ will effectively be the convolution of $p(\Lambda)$ with a selection function whose form is only loosely defined and which will depend on complex questions of measure  and anthropic selection \cite{DeSimone:2008bq}.  

In this paper we focus on simple Gaussian random landscapes, and find that the relative numbers of extrema and saddles are roughly but not exactly binomial.  The analysis here is effectively an $N$-dimensional generalization of the approach taken in the now classic treatment of the theory of  fluctuations in the  density profile of the early universe of Bond, Bardeen, Kaiser and Szalay  \cite{Bardeen:1985tr}, or BBKS. We also provide arguments that this may be due to topological constraints and hence applicable beyond random Gaussian potentials.

A large set of cosmological parameters is associated with the properties of the landscape and we begin with a survey of these observables and the analogous properties of $\poten(\bar\field)$ in Section \ref{sec:observables}.   In Section \ref{sec:Topology} we provide several topological hints on the relative number of different types of stationary points and in Section \ref{sec:threeFieldCounting} we provide the formalism for calculating relative number of different types of stationary points. In Section \ref{sec:largerN} we present  the results of our calculation for  up to 100 fields and compare our results with the large-N limit results obtained in \cite{Bray:2007tf}. Finally we conclude in Section \ref{sec:conclusion}.

\section{Cosmological Properties and Landscape Architecture \label{sec:observables}}

As is now well-established, an inflationary phase in the early universe can resolve the initial conditions problems faced by simple models of the hot big bang \cite{Guth:1980zm,Linde:1981mu,Albrecht:1982wi}. However there is no unique mechanism to drive the accelerated expansion associated with inflation \cite{Ade:2015lrj}, and several hundred different models have been proposed and examined \cite{Martin:2013tda}. Typically these models are specified by  the effective potential of the inflaton field(s). If it is assumed that the inflationary potential is contained somewhere in the overall landscape,  the ``typical'' inflationary mechanism in a landscape will be associated with the expectation values of derivatives of the fields on candidate inflationary trajectories. 

Let us begin by surveying the range of observables that may be associated with a landscape potential, and the properties of the potential that determine them:
\begin{itemize}
\item  The distribution of  vacuum energies,  $p(\Lambda)$, in the landscape (see e.g. Ref.~\cite{Weinberg:2000qm}).  In a given pocket \cite{Guth:2000ka} this corresponds to local minima of the landscape, with the value of the vacuum energy / dark energy density $\Lambda = \poten(\bar \field)$ at each minimum. 
\item  The stability of the vacuum as a function of $\Lambda$ \cite{Lee:1974ma,HAWKING198235}. The  local vacuum in a landscape potential is typically   metastable due to bubble nucleation  via quantum tunneling \cite{Coleman:1977py,Callan:1977pt, Coleman:1980aw,Greene:2013ida,Dine:2015ioa,Masoumi:2016eqo} and  will be  unstable if there is a noticeable probability of decay within cosmologically  relevant time scales. 
\item Bubble collision \cite{Chang:2008gj,Aguirre:2009ug,Feeney:2010jj}. Collision rates depend on the nucleation rate which is a function of $V(\bar \phi)$ at local minima and the surrounding barrier heights.%
\item  The likelihood of slow roll inflation. Slow roll inflation requires  sufficiently long, flat ``plateaus" or ``valleys'' in $\poten(\bar \field)$, and that $V,_{i}$ and $V,_{ij}$  in downhill directions in the potential are  parametrically small. 
\item Primordial perturbation spectrum. Expectation values for the spectral index $n_s$ and tensor to scalar ratio $r$ (and  correlations between them) are derived from the expected values of  $\poten,_{ij}$ and $\poten,_{i}$ along  inflationary trajectories. 
\end{itemize} 

While the landscape may be almost arbitrarily complex, its overall form may be motivated by a handful of fundamental physical principles. The heart of  this proposal is to use these principles to identify the overall {\em architecture\/} of the landscape. Such an architecture can lead to specification of a detailed probability distribution from which $V(\bar \phi)$, the potential energy function of the landscape, can be drawn. 
 
To illustrate this approach, the simplest landscape architecture  we can imagine consists of $N$ fields,  $\field_1, \field_2, \ldots, \field_N$ with  self-interaction potentials $\poten_i(\field_i)$ and no mutual interactions, so that the landscape potential $\poten$ is
\bel{indepenSum}
	\poten(\field) = \sum_{i=1}^N \poten_i({\field_i})~,
\ee 
which each $V_i$ is to be chosen probabilistically. Even without specifying the probability distribution for each $V_i$ we know that the number of maxima and minima cannot differ by more than one, which would be negligible if the number of stationary points is large. If the $\poten_i(\field_i)$ are periodic, then  the number of maxima and minima must match exactly. For a given stationary point, the probability that $k$ of the eigenvalues of the corresponding Hessian are positive is  exactly  \cite{Aazami:2005jf}
\bel{indProb}
	P_k= 2^{-N}{N \choose k}~. 
\ee
We thus immediately deduce that the ratio of the number of minima to stationary points is $1:2^N$ for a landscape that consists purely of uncoupled, self-interacting fields. The central limit theorem implies that, for sufficiently large $N$, the distribution of vacuum energies would be a Gaussian \cite{Masoumi:2016eqo}.

Conversely assuming that all fields have similar mutual- and self-interactions and that the overall landscape potential is a combination of many individual terms motivates a landscape architecture that consists  of  a Gaussian random function,
\bel{potentialG}
	\poten(\bar \field)=  \pot(\bar \field),
\ee
where $\bar \field$ denotes the $N$ scalar fields and $\pot(\bar \field)$ is a Gaussian random function. If we stipulate that there are no preferred directions or positions in the landscape then the correlation function will naturally be  rotationally invariant: 
\bel{Correlation}
	\langle  \pot({ \bar \field_1} )   \pot({ \bar \field_2} ) \rangle  =N  f \left( \frac{( \bar \field_1 -  \bar \field_2)^2}{N} \right) ~.
\ee
Likewise, given that we aim to investigate the implications of a given landscape architecture  for the cosmological constant problem, it is natural to stipulate that the mean of $U$ is zero.  A more sophisticated architecture  arises from assuming that $\pot(\bar \field)$ is associated with new physics at a very high but sub-Planckian scale $M$ (e.g. string or GUT-scale physics) and that Planck-scale operators induce an extra correlation at large VEVs, so that 
\bel{potential}
	\poten(\bar \field)= \poten_0 + \frac12 m^2  \bar \field \cdot  \bar \field+ \pot(\bar \field),
\ee
where $\pot$ is again a Gaussian random function and we have also added an arbitrary offset $V_0$ for generality.  If  $|\bar \field|$ can approach Planckian values, then $m^2 M_p^2 \sim |\pot| \sim M^4$ at the ``edge'' of the landscape, or
\bel{massterm}
	m = M {M \over M_{\rm P}}~.
\ee
so that 
\be
	\langle \pot(\bar \field) \rangle=0 ~, \qquad  \langle \pot^2(\bar \field) \rangle= {\cal{O}}(M^8)~.
\ee 
By hypothesis, $M$ is significantly smaller than the Planck mass and we also require the typical correlation length of $\pot(\bar \field)$ to be much less than $M_{\rm P}$. In this case we will naturally expect $\pot$ to contain a large number of extrema.

In this paper, our goal is to understand the properties of the critical points in this landscape. Any specific critical point of a function $V$ is characterized by the Hessian
 \bel{Hessian}
	 \hs_{ij}={\partial^2 V \over \partial \field_i \partial \field_j}~.
 \ee
In the absence of any other information, it may be tempting to assume that the Hessian is a symmetric random matrix  \cite{Aazami:2005jf}. Denoting the (ordered) eigenvalues of $\hs(\field_{\rm st})$ by $\lambda_1 \ge \lambda_2 \ge \ldots \ge \lambda_N$ one might naively assume that because each eigenvalue is equally likely to be positive  or negative the likelihood that $\field_{\rm st}$ is a vacuum (local minimum) is  $(1/2)^N$. However, the joint probability distribution for the $\lambda_i$ is
 \bel{eigDist1}
 	P(\lambda_1, \ldots, \lambda_N) \propto \prod_{i<j} (\lambda_i-\lambda_j)~. 
 \ee
and the likelihood that all the eigenvalues are positive scales as  $e^{-\alpha N^2}$ \cite{Aazami:2005jf,Dean:2006wk}. However, as we will describe in detail below, the Hessians of random functions are not random matrices with  independent and identically distributed components. As we   show below, the expected fraction of minima is much closer to the binomial form $2^{-N}$ than $e^{-\alpha N^2}$.

\section{Topology and Morse Theory}
\label{sec:Topology}

One way to approach the landscape is to assume that the Hessian matrices are random, symmetric matrices drawn from the GOE. If we wish to apply the standard results of random matrix theory to the Hessians derived from a given random function, it is necessary for the individual elements of the Hessian to be drawn from independent and identical distributions. However, the Hessians of a random function are not uncorrelated; consider two elements of the Hessian; $\hs_{ij}$ and $\hs_{kl}$; given than mixed derivatives commute, working from the definition of the Hessian it follows that 
\bel{HDerCorrelation}
	\partial_k\partial_l \hs_{ij}= \partial_i \partial_j \hs_{kl}~.
\ee
Consequently, the elements of the Hessian matrices derived from a specified function are not independent of one another.  

\begin{figure}[tbp] 
   \centering
   \includegraphics[width=3in]{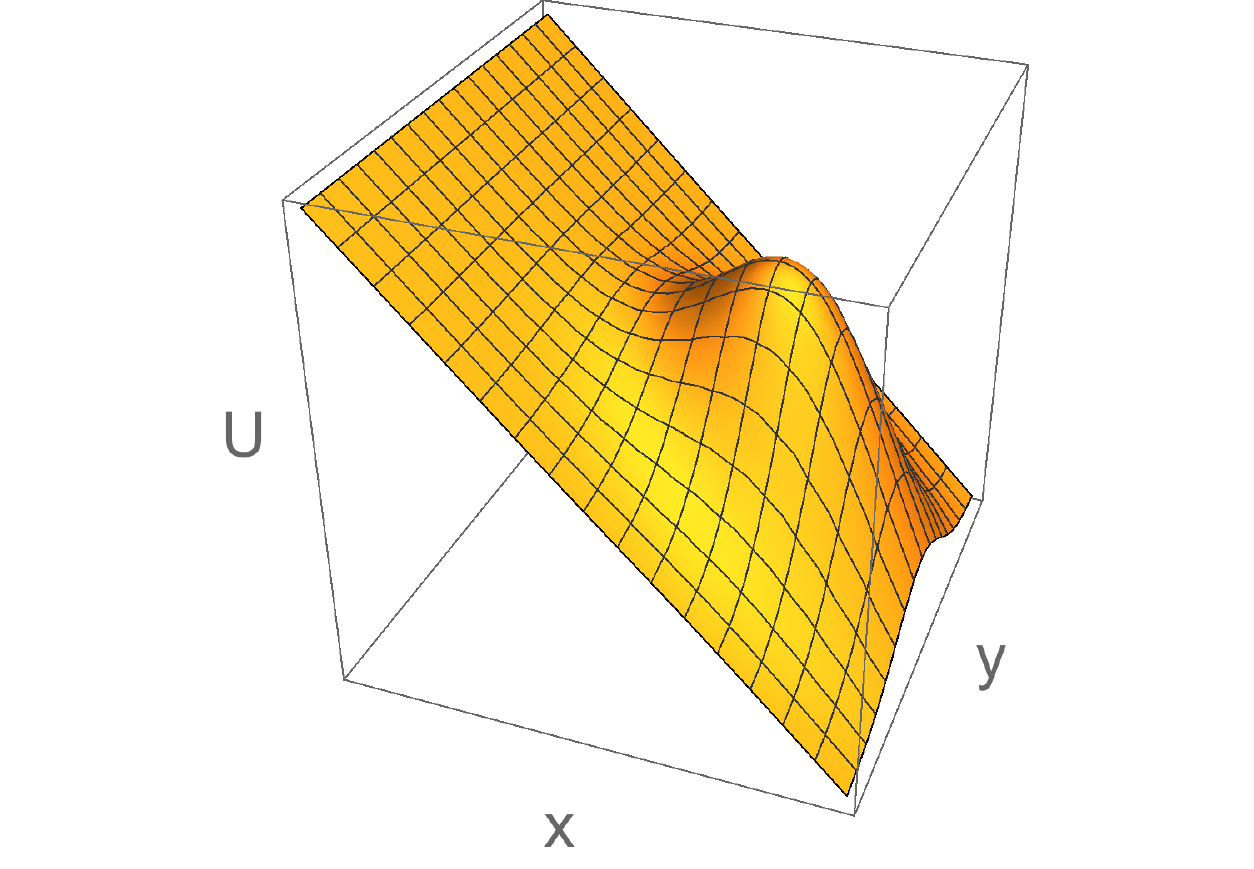} 
   \caption{One can add a maximum and a saddle point to a straight slope without creating any new minimum illustrating that \eqref{Morse2} is valid only statistically and not for all individual realizations of the potential.}
   \label{fig:AddMax}
\end{figure}

In the next sections we will directly compute the fraction of extremal points of an $N$-dimensional Gaussian random function which are actual minima. However, we can also gain significant insight from global, topological arguments based on Morse Theory.  To start with, consider a two-dimensional periodic landscape. Let's $N_{\rm max}, N_{\rm min}$ and $N_{\rm saddle}$ denote the number of maxima, minima and saddle points of this function. From Morse theory we know  
\bel{Morse1}
	N_{\rm min}-N_{\rm saddle}+ N_{\rm max}= \chi_{\rm torus}=0~,
\ee
where $\chi_{\rm torus}$ is the Euler characteristic of the torus. This result holds for any function on torus and it must be satisfied case by case and not an average.  For random Gaussian functions the symmetry of $U\rightarrow - U$ ensures  $N_{\rm max}=N_{\rm min}$ which combined with \eqref{Morse1} gives 
\bel{Morse2}
N_{\rm saddle}= 2 N_{\rm max}= 2 N_{\rm min}
\ee
We calculated the same quantities using a GOE  
\bel{randomGaussianPred}
	N_{\rm saddle}=2 \left(1+\sqrt{2}\right) N_{\rm min}\approx 4.82 N_{\rm min}~.
\ee
 \begin{figure}[tbp] 
   \centering
   \includegraphics[width=4in]{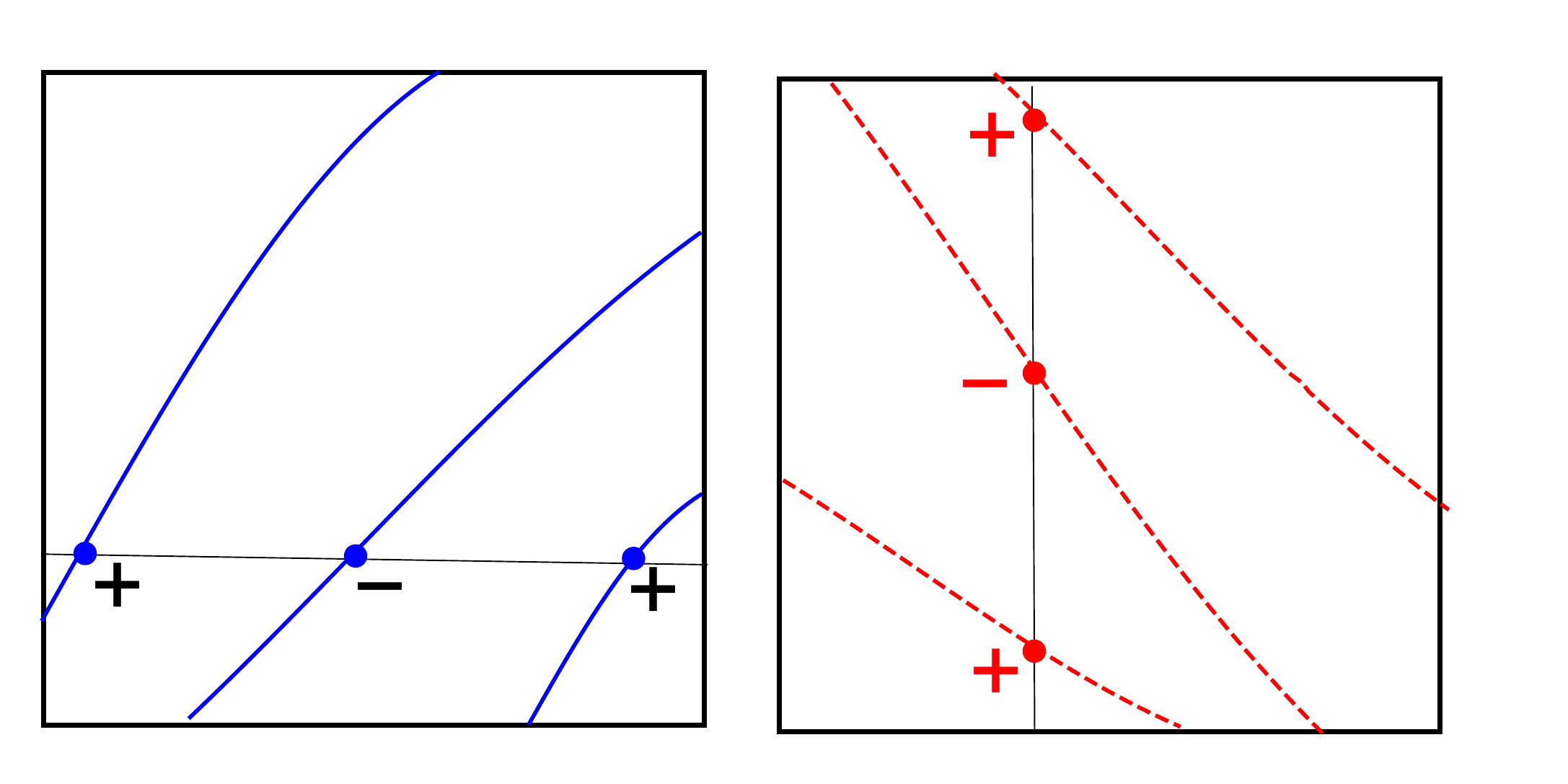} 
    \includegraphics[width=2.5in]{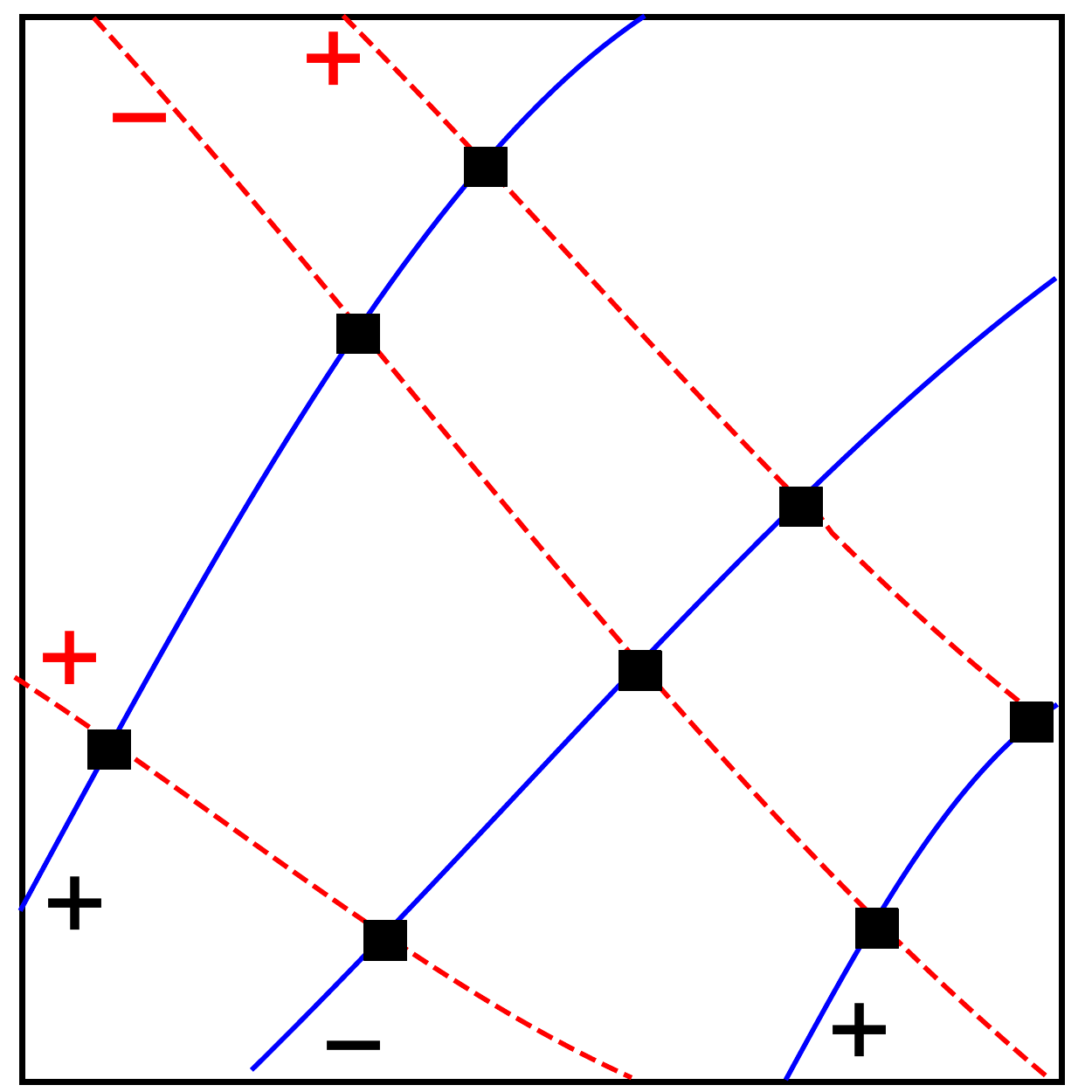} 
   \caption{Top left and right, the contours for $\partial \poten/\partial \field_1=0$ and $\partial \poten/\partial \field_2=0$, bottom the two graphs super-imposed. Sign of $\partial^2 U/\partial\field_1^2$ and $\partial^2 U/\partial \field_2^2$ are shown by black and red plus and minus signs. It is clear that the stationary points of the potential which are located at the black filled squares have alternating signs for diagonal elements of the Hessian $\hs$.}
   \label{fig:heuristic}
\end{figure}
Notice that to derive \eqref{Morse2} we used an average symmetry of $U\rightarrow -U$ which does not hold in general. Therefore,  \eqref{Morse2} will not be valid for each realization of the potential. A counter example is shown in Figure~\ref{fig:AddMax}.

This result generalizes to higher numbers of dimensions. If $N_i$ is the number of stationary points with $i$ negative eigenvalues in the Hessian, we get 
\bel{MorseGeneral}
	\sum_{i=0}^N (-1)^i N_i = \chi({\cal M})~,
\ee
where $\chi({\cal M})$ is the Euler characteristic of the manifold $\cal M$. It is easy to show that the random Gaussian matrices in even number of dimensions do not satisfy this requirement. 

We can  give a heuristic argument  that there are important correlations between  elements of the Hessian at adjacent stationary points.  This argument is illustrated in Fig.\ref{fig:heuristic}.  A stationary point of the potential is given by $\nabla U(\field)=0$.  Let's first look at the hyper-surfaces  $\partial U/\partial \field_1=0$ which is shown on the top left panel. Along the straight horizontal line, the potential is a one-dimensional function, and therefore the sign of the second derivative along the line  must change at the stationary points (blue dots).  Consequently,  the corresponding Hessian element $\hs_{11}$ must change sign along this line. Moreover, the sign of $\hs_{11}$ must be constant on each hypersurface. It can only change if there are horizontal lines where $\hs_{11}$ vanishes  at all blue dots, which is highly unlikely, or if the topology of the contours change. Similarly for the hypersurfaces $\partial U/\partial \field_2=0$  shown in the top right panel must have alternating signs for $\hs_{22}$. The stationary points  lie on the  intersections of these contours which are shown as black filled squares in the lower panel. Therefore the sign structure of the Hessian diagonal elements $\hs_{ii}$ is very similar to the one for sum of independent potentials presented in \eqref{indepenSum}. Moreover, by a change of coordinate which mixes the diagonal and non diagonal elements, there will be correlations between the non-diagonal elements of the adjacent stationary points. This strongly suggests that the Hessian cannot be modeled with a random Gaussian matrix and there are always important correlation between the Hessian at different stationary points. This is only a heuristic argument and by no means a proof but it seems to be a generic situation and should hold statistically and our results in the later sections suggest that this argument is plausible. Moreover, this statement applies to any function not just random functions.

\section{Counting stationary points of random Gaussian fields} 
\label{sec:threeFieldCounting}
Here we start from a random Gaussian potential as defined in \eqref{Correlation}. Without making any assumption about the Hessian, we directly calculate the number of different types of stationary points by closely following the methods of BBKS \cite{Bardeen:1985tr}. Because these potentials are translationally invariant we  end up with the number density of these stationary points. Let's first rewrite \eqref{Correlation} in Fourier space and define the power spectrum $P(k)$
\beal{Correlation2}
	\langle \pot(\bar \fn_1)\pot(\bar \fn_2)\rangle &=&\frac1{(2\pi)^N} \int d^Nk\; P(k) e^{i\bar k \cdot (\bar \fn_1 - \bar \fn_2)}~.
\eea
Because $\pot$ is a Gaussian random field, all the odd moments of the distribution vanish. We use the following  notation for the even moments and gradient :
\bel{moments}
	\sigma^2_{n} = \frac1{(2\pi)^N}\int d^N k (k^2)^n P(k)~, \quad \gra_i = \frac{\partial \pot}{\partial \fn_i}~.
\ee
We denote the eigenvalues of the $\hs$, the  Hessian, as $\lambda_1, \ldots, \lambda_N$ and to specify them unambiguously we choose  $\lambda_1 \ge \lambda_2 \ge \ldots \ge\lambda_N$. We use a variant of the Kac formula \cite{KacRice} to compute the number of stationary points with $P$ positive eigenvalues ($P\ge1$) in a region,
\bel{density1}
	N_P= \int d^N\fn \; \delta^N(\gra_i) \left|{\rm det} (\hs_{ij})\right| \theta_H(\lambda_{P}),
\ee
where $\delta^N$ is the $N$-dimensional Dirac delta and $\theta_H$ is the Heaviside step function. The ability of this expression to ``count'' extrema follows by noting that it is nonzero only at stationary points and at those points the Jacobian is cancelled by the change of variables in $\delta$ function, and its overall value would be $\pm1$ depending on the number of negative eigenvalues. Similarly, the number of maxima is given by
\bel{maxDensity}
	N_0= \int d^Nx \; \delta^N(\pot_{,i}) {\rm det} (\pot_{,ij}) \theta_H(-\lambda_1).
\ee
The functions  $\pot$, $\gra_i$ and $\hs_{ij}$ are Gaussian variables with zero mean and all  we  need is their standard deviation. It is now easy to show that the only nonzero two-point functions are 
\beal{correlations2}
	\expec{\pot(\bar \field) \pot(\bar \field)}&=&\sigma^2_0~, \nn
	\expec{\gra_i(\bar \field)\gra_j(\bar \field)}&=& \frac{1}N \delta_{ij}\sigma^2_1~, \nn
	\expec{\pot(\bar \field)\hs_{ij}(\bar \field)}&=& -\frac{1}N \delta_{ij}\sigma^2_1~, \nn
	\expec{\hs_{ij}(\bar \field)\hs_{kl}(\bar \field)}&=& \frac{ \sigma^2_2 \( \delta_{ij} \delta_{kl}+  \delta_{il} \delta_{jk}+  \delta_{ik} \delta_{jl}\)}{N(N+2)} . 
\eea
To calculate the total number of stationary points of a specific type from  \eqref{density1} we define the following vector 
\beal{alphdaDef}
	\alpha_i&=&\{\pot, \gra_{1}, \gra_2, \ldots, \gra_{N},\hs_{11}, \hs_{22},\ldots, \hs_{NN}, \hs_{N-1,N} ,\hs_{N-2,N}, \nn
	&&   \ldots, \hs_{1N},\hs_{N-2, N-1}, \ldots, \hs_{1,N-1}, \ldots, \hs_{12}\}~. \label{alpha2}
\eea
This vector has $\frac12 N(N+3)+1$ elements and we define  
\bel{mMatrix}
	M_{ij}\equiv\expec{ \alpha_i \alpha_j}~, \; \text{and  }\; K_{ij}\equiv M^{-1}~.
 \ee 
Because $\alpha_i$ have  normal distribution, their joint probability distribution is given by
\bel{alphaProb}
	p(\alpha_i)= \frac{\sqrt{{\rm det} K}}{(2\pi)^{N(N+3)/4} } e^{-Q}\; \; \text{where} \;\;Q= \frac12 \alpha K\alpha~.
\ee
It is easy to check that \eqref{alphaProb}  gives the correlation functions given in \eqref{correlations2}. In order to get the total number of a given type of stationary point we evaluate the integral given in \eqref{density1} and \eqref{maxDensity} using this probability distribution. For example, for $P>0$, we get
\begin{eqnarray}
	N_P&=& \int d^Nx \; \prod_{i=1}^{1+N(N+3)/2} \!\!\!\!\!\!\!\!\!d\alpha_i \;\;\delta^N(\gra_i)  {\rm det} (\hs_{ij}) \theta_H(\lambda_{P})  p(\alpha_i).\label{density1}
\end{eqnarray}
The most general real symmetric matrix $\hs_{ij}$ can be written as 
\bel{eulerCoord}
	\hs= R^T \left( \begin{array}{ccc}
	\lambda_1 &   \ldots & 0 \\
	\vdots & \ddots & \vdots \\
	0 &\ldots & \lambda_N
	\end{array}\right) R~,
\ee
where $R$ is a rotation matrix that diagonalizes it. One way to evaluate these integrals is to parametrize the Hessian as in \eqref{eulerCoord} and write the matrix $R$ in terms of $N(N-1)/2$ Euler angles. Adding the $N$ eigenvalues, we recover the $N(N+1)/2$ independent parameters which  specify a symmetric matrix. This approach is tractable for $N=2$ and 3  and we present explicit results in Appendices \ref{sec:NEq2} and \ref{sec:Neq3}. In particular, for $N=2$ the ratio of Hessians with zero, one and two positive eigenvalues is $1:2:1$ while for $N=3$ the analogous ratio is $1:3.05:3.05:1$, showing that the number of relative number of minima and saddles is close to (and exact, for $N=2$) the result for uncoupled potentials.  

\begin{figure}[tb]
 \includegraphics[width=3in]{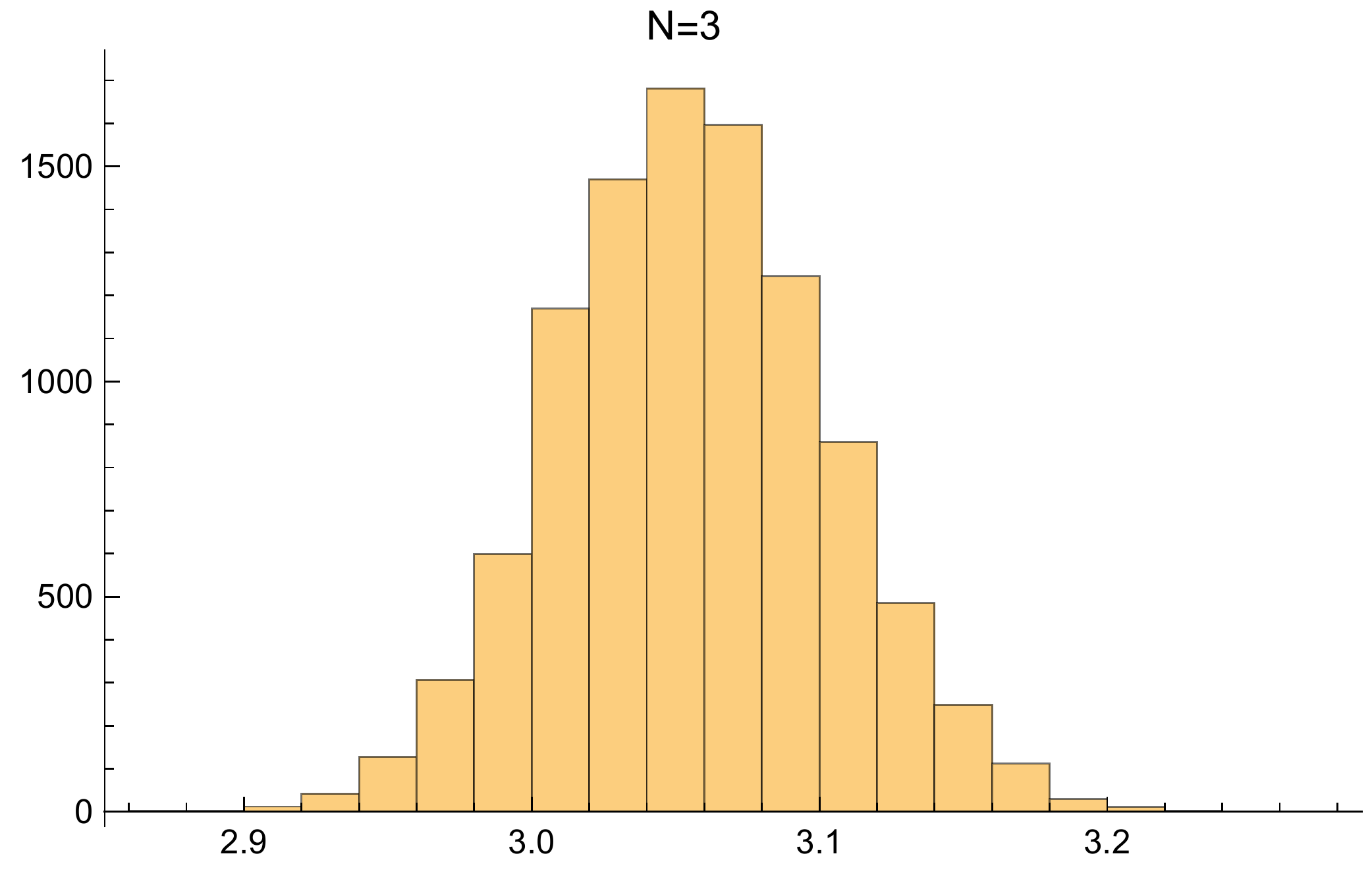} 
\caption{Ratio of saddles to minima for a 3-d function with spherical cutoff; clearly centered around 3.05. \label{fig:numeric}}
\end{figure}

For comparison, we computed the expectations for the eigenvalue distribution for matrices drawn from the GOE, by integrating over the relevant measure. For $N=3$ we get these ratios
\be
	n_{\rm rel}=\left\{1,1+\frac{8}{\sqrt{2} \pi -4},1+\frac{8}{\sqrt{2} \pi -4},1\right\}=\{1,19.06,19.06,1\}~.
\ee
Separately, we directly generated numerical realizations of random functions, by generating  realizations of band-limited Fourier sums. We looked at two cases, a spherical cutoff including all modes with $|\bar k|$ less than a fixed cutoff and a Gaussian, scale invariant power spectrum.\footnote{We also analyzed a ``Cartesian'' landscape in which $k_1$, $k_2$ and $k_3$ are summed over, up to a fixed cutoff. This function is anisotropic in $k$ and  does satisfy the criteria that define the random functions analyzed here. The corresponding ratios were $1:3:3:1$, which is to be expected since there is more high-frequency power in  directions where $|k_1| \sim |k_2| \sim  |k_3|$ as the Hessian matrices in a diagonal basis are dominated by the diagonal terms.}   Averaging over multiple realizations we find:
\bea
	n_{\rm rel}&=&\{1.,3.056,3.057,1.0008\}~, \qquad  \text{Spherical}~,\nn
	n_{\rm rel}&=&\{1.,3.055,3.055,1.0003\}~, \qquad \text{Gaussian}~. 
\eea
The distribution of the relative number of the saddle points to minima and maxima for an scale invariant power spectrum for the $N=3$ case is shown in Figure~\ref{fig:numeric}.

\section{Density of stationary points  for general $N$}
\label{sec:largerN}

For  $N>3$ it turns out to be more convenient to use the techniques of BBKS   \cite{Bardeen:1985tr}, generalized to $N$ dimensions. We will need the integration measure which is the Jacobian of the transformation from the Hessian $\hs_{ij}$ to the Euler angles and eigenvalues. We obtain this by noting that if we have a metric (inner product) on this space, the square root of its determinant  gives the integration measure. Following the overall approach of BBKS, we use the following  inner product on the space of symmetric $N\times N$ matrices  
\bel{inner}
	S_1 \cdot S_2 = {\rm Tr} (S_1 S_2)~.
\ee
It is obvious that $\Tr \[S_1(S_2+ \alpha S_3)\]= \Tr (S_1) + \alpha \Tr (S_2)$, $\Tr (S_1 S_2) = \Tr (S_2 S_2)$ and $\Tr (S_1 S_1) = \sum \lambda_i^2 \ge0$  and hence this is a valid metric. If a matrix $S$ can be diagonalized by an orthogonal matrix $R$, the matrix  $S+ dS$ will be diagonalized by   $R+dR$  and if $R^\dagger S R= {\rm diag} (\lambda_1, \lambda_2, \ldots, \lambda_N)$, the eigenvalues would change by $d\lambda$. Using $R^\dagger R=I$
\bel{InverseDer}
R^\dagger dR + d R^\dagger R=0~,
\ee
which translates to $dR^\dagger= -R^\dagger dR R^\dagger$. Therefore\footnote{We have corrected a typo in Appendix B of BBKS. }
\beal{BBKSAppeB1}
	dS = R^\dagger d\lambda R +R^\dagger \lambda dR +  dR^\dagger \lambda R  &=&  R^\dagger d\lambda R +R^\dagger \lambda dR -  R^\dagger dR R^\dagger \lambda R \nn&=& R^{\dagger} \(d\lambda + [\lambda, dR R^\dagger] \) R~.
\eea
From here we get 
\be
	ds^2 = \Tr (dS^2 ) = \Tr\( d\lambda^2 + [\lambda, dR R^\dagger]^2 + 2 d\lambda [\lambda, dR R^\dagger]\)~.
\ee
Because $d\lambda$ and $\lambda$ are both diagonal  they commute and the last term vanishes, leaving
\bel{metric2}
	ds^2 = \Tr (dS^2 ) = \Tr\( d\lambda^2 + [\lambda, dR R^\dagger]^2 \)~.
\ee
Noticing that \eqref{InverseDer} shows that $R^\dagger dR$ is an antisymmetric matrix  the trace in \eqref{metric2} is given by
\bel{metric3}
	ds^2= \sum (d\lambda_i)^2 + \sum_{i\neq j} (\lambda_i-\lambda_j)^2 \omega_{ij}^2~
\ee
where $\omega_{ij} = (dR R^{\dagger})_{ij}$ depends only on elements of the rotation group which are given by Euler angles. There are $N+ N(N-1)/2=N (N+1)/2$ orthogonal vectors in \eqref{metric3} and hence these form the orthonormal set we were looking for. Therefore the volume element is 
\begin{eqnarray}
	d {\rm Vol} &=& \prod_{i\le j}d \hs_{ij} = \prod_{i=1}^N d\lambda_i \prod_{i\neq j} (\lambda_i-\lambda_j) \Omega(\alpha_k) \prod_{k=1}^{\frac12 N (N-1)}d\alpha_k =F(\alpha_i)\prod_{i=1}^N d\lambda_i \prod_{i\neq j} (\lambda_i-\lambda_j)~.
\end{eqnarray}
Because of spherical symmetry, the only dependence on the Euler angles $\alpha_i$ would be through a normalization factor which is irrelevant for the calculation of the relative number of different types of stationary points. To evaluate $Q$  in \eqref{alphaProb} we find the inverse of the  $M$  in \eqref{mMatrix} and we later  evaluate $Q$ on the surface $\gra_i=0$. The nonzero elements of $K$ are 
\beal{Kelems}
	K_{\pot\pot}&=&\frac{\sigma _2^2}{\sigma _0^2 \sigma _2^2-\sigma _1^4} ~, \nn 	
	K_{\gra_i \gra_j} &=& \frac N{\sigma_1^2}~, \nn
	K_{\pot, \hs_{ij}}&=&\frac{\sigma _1^2}{\sigma _0^2 \sigma _2^2-\sigma _1^4}~,\nn
	K_{\hs_{ij}\hs_{kl}}&=&\frac{\sigma _1^2}{\sigma _0^2 \sigma _2^2-\sigma _1^4}\delta_{ij}\delta_{kl}+\frac{N(N+2)}{2\sigma_2^2}\(2\delta_{ik}\delta_{jl}-\delta_{il}\delta_{jk} \)~.
\eea
We evaluate $Q$ on the surface $\gra_i=0$
 \beal{Qdef}
	Q &=&\frac1{\sigma_0^2 \sigma_2^2 -\sigma_1^4} \(\frac12 \sigma_2^2 \pot^2 + \sigma_1^2 \pot \Tr \hs  + \frac{(N+2)\sigma_1^4- N \sigma_0^2 \sigma_2^2}{4\sigma_2^2}(\Tr \hs)^2  \)+\frac{N(N+2)}{4 \sigma_2^2} \Tr \hs^2~.
\eea
From here we can easily show that at a constant $\pot$ 
\beal{expectations3}
	\expec{\hs_{ij}}_\pot&=&-\frac{\sigma_1^2}{N \sigma_0^2} \pot\delta_{ij}~,\nn 
	\expec{\hs_{ij}\hs_{kl}}_\pot&=&\(-\frac{\sigma_1^4}{N^2 \sigma_0^2}+\frac{\sigma_1^4 \pot ^2}{N^2 \sigma_0^4}\) \delta_{ij}\delta_{kl} + {\sigma_2^2 \over N(N+2)} (\delta_{ij}\delta_{kl} +\delta_{ik}\delta_{jl} +  \delta_{il}\delta_{jk}) ~.
\eea
Similar results for a very specific power spectrum was obtained in \cite{Bachlechner:2014rqa} which looked at the distribution of minima for a power spectrum given by 
\beal{correlationThom}
	\langle \pot(\bar \fn_1)\pot(\bar \fn_2)\rangle &\propto& e^{-|\field_1 - \field_2|^2/\Lambda_h^2}~,
\eea
where $\Lambda_h$ is the correlation length. It is easy to check that in this special case the term which has $\({(N+2)\sigma_1^4- N \sigma_0^2 \sigma_2^2}\) \Tr \hs^2$ in \eqref{Qdef} vanishes and then the Hessian can be thought as the sum of a GOE plus a coefficient times the identity matrix. However, as it is clear from \eqref{Qdef} this is not true in general. To see it more clearly, if one could write $\hs= W + p(\pot) I$ where $W$ is chosen form a GOE, then the term proportional to $(\Tr \hs)^2$ in \eqref{Qdef} would not appear. Therefore, except for very special cases, the Hessian is not the sum of a matrix from GOE and a matrix proportional to identity.  Interestingly, the second equation in \eqref{expectations3}  shows that even when the potential is zero the  elements of the Hessian are not generically drawn from a random orthogonal matrix. Because in this paper we are only interested in the number of stationary points and not their energy we integrate over the potential to get the distribution of $\hs$. The probability distribution of the Hessians is give by
\bel{newProb}
	P(\hs)\propto e^{-\tilde Q}
\ee
where 
\bel{Qgeneral}
	\tilde Q= \frac N{4\sigma_2^2} \[ (N+2)\sum \Tr \hs^2  -(\Tr \hs)^2\]~.
\ee
The last ingredient we would need for an explicit calculation of the number of stationary points of a given kind is $\det M$ which is given by
\begin{eqnarray}
	\det M&=&{2^{N-1}\over (N+2)} \(\frac{\sigma_1^2}{N}\)^N  \(\frac{\sigma_2^2}{N(N+2)}\)^{N(N+1)/2} = G(N) \sigma_1^{2N} \sigma_2^{N(N+1)}~.
\end{eqnarray}
The density of stationary points of a given type (using the appropriate $\theta$ function) is given by
\beal{densityGeneral}
	n_i &=&\frac{1}{(2\pi)^{N(N+3)/2} \sqrt{\det M}}  \quad  \int \prod d\lambda_i \prod \lambda_i \prod_{i>j}(\lambda_i-\lambda_j) e^{-\tilde Q} \theta_H \int \prod d{\alpha_i} F(\alpha_i)\nn
	&=&\frac{F_N}{(2\pi)^{N(N+3)/2} \sqrt{\det M}} \quad \int \prod d\lambda_i \prod \lambda_i \prod_{i>j}(\lambda_i-\lambda_j) e^{-\tilde Q} \theta_H ~,
\eea
where $F(N)$ is the normalization factor we  get by integrating over all Euler angles which we can ignore in what follows.  The number of stationary points depends on a geometrical factor from Euler angles and  the only dependence on the power spectrum is  through
\be
	n_i \propto \frac{ \sigma _2^N}{ \sigma _1^N}~.
\ee
This dependence does not change the relative number of stationary points. It is fascinating to notice that the relative number of stationary points is independent of the power spectrum and depends only on the ratios of integrals over the $\lambda_i$.  We cannot evaluate this integral analytically for arbitrary $N$,  but we can make substantial numerical progress.  For  $N\le 9$ we used the implementation of VEGAS \cite{Vegas},  an adaptive Monte Carlo method, in the GNU Scientific Library (GSL) \cite{GSL} with sample sizes of $10^7$ to $10^9$ points to get percent-level precision.  However, for $N\ge10$, the GSL implementation failed with underflow errors but a purpose-written implementation of Metropolis-Hastings \cite{Metropolis, Hasting} algorithm allowed us to reliably evaluate these integrals for up to $N=100$. Detailed results for $N=50$ are presented in Appendix \ref{sec:NEq50}.\footnote{This code was implemented in Power-BASIC  and run on several desktop machines.}  Finally, we verified this approach for $N\le 50$ using the PolyChord \cite{Handley:2015fda} -- this package is designed to calculate Bayesian evidence which (for a suitable choice of likelihood function) is mathematically equivalent to the problem faced here. 

We express  our densities relative to $n_0$, the density of minima for each value of $N$, and write $n_i=n_0 \beta_i$.  We show the results of $n_0$ for $N\le 10$ in Table \ref{table:Smalln0} and the corresponding  $\beta_i$'s in Table \ref{table:SmallNbetai}. The values computed and presented in Table \ref{table:Smalln0} contain an overall geometric factor that scales as the relative volume of the $N$-sphere which accounts for the very small numerical values, and this term cancels from the ratios found in Table \ref{table:Smalln0}.    The $\beta_i$'s are relatively  close to a binomial distribution, which is the exact result for landscape that is a sum of independent potentials in \eqref{indepenSum}. For contrast, the  $\beta_i$  that would be expected if  the distribution of extrema was controlled by the relatives numbers of same-sign eigenvalues in matrices drawn from the GOE are shown Table.\ref{table:randomHessian}. These numbers are dramatically different from both the explicit results we obtained for a Gaussian random function and the pure binomial distribution. 
\begin{table*}
\begin{tabular}{|c|c|c|c|c|c|c|}
\hline 
$N$ & 4 & 5 & 6 & 7 & 8 & 9   \\ \hline 
$n_0 \sigma_1^N/(F_N\sigma_2^N)$& $3.2\times 10^{-11}$ & $1.5 \times 10^{-15}$ & $2.1 \times 10^{-20}$
& $1.1 \times 10^{-25}$ &$2.3 \times 10^{-31} $  &$2.0 \times 10^{-37} $  \\ \hline
\end{tabular}
\caption{Numerical values for the density of minima for $N=4,\ldots,9$~. }
\label{table:Smalln0}
\end{table*}

\begin{table*}
\begin{tabular}{|c|c|c|c|c|c|c|c|c|c|c|c|c|}
\hline
$N$ &  $\beta_0$& $\beta_1$& $\beta_2$& $\beta_3$& $\beta_4$& $\beta_5$  & $\beta_6$& $\beta_7$& $\beta_8$&$\beta_9$  \\ \hline
4 & 1 & 4.04 & 6.08 & 4.04 & 1.00 &&&&& \\ \hline
5 & 1  &5.36    & 11.08& 11.08 & 5.36 &1.00 &&&&\\ \hline
6 & 1& 6.62& 17.45& 23.68& 17.45& 6.614& 1.00&&& \\ \hline
7 & 1& 8.09& 26.2& 45.3& 45.3& 26.2& 8.1& 1.02&& \\ \hline
8 & 1& 9.28& 36.0& 76.& 96.6& 76.& 36.& 9.28&  1.00 &\\ \hline
9 & 1& 10.9& 49.1& 123.& 192.& 192.& 123.& 49.3& 10.9& 1.00 \\ \hline

\end{tabular}
\caption{Numerical results of $\beta_i$ for $N>3$. We see this is very close to the ratios we get for independent potentials described in \eqref{indepenSum}. These numbers are accurate within one percent error. }
\label{table:SmallNbetai}
\end{table*}
\begin{table*}
\begin{tabular}{|c|c|c|c|c|c|c|c|c|}
\hline
$N$ &  $\beta_0$& $\beta_1$& $\beta_2$& $\beta_3$& $\beta_4$& $\beta_5$  \\ \hline
2 &  1 & 4.83& 1&&& \\\hline
3  & 1 & 19.0 &19.0 &1&&\\ \hline
4 & 1 & 72.0  & 261.1 & 72.0 & 1& \\ \hline 
5& 1 & 268.7&3299.& 3299.& 268.7&1\\ \hline
\end{tabular}
\caption{The values of $\beta_i$'s for a GOE. We see that this is not a good model for the distribution of stationary points of the landscape and for even number of dimensions it also does not satisfy the criteria from Morse theory.}
\label{table:randomHessian}
\end{table*}

\begin{figure}[htbp] 
   \centering 
      \includegraphics[width=4in]{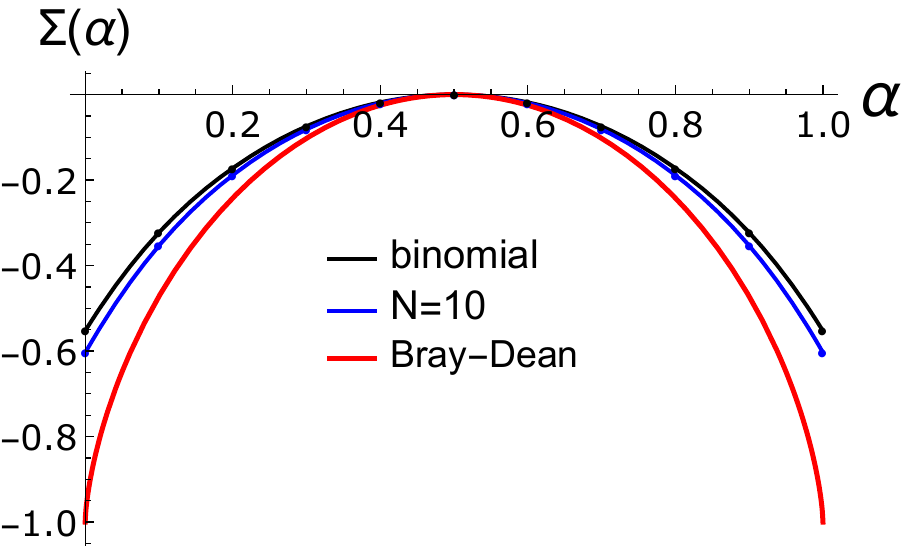} \\
   \includegraphics[width=4in]{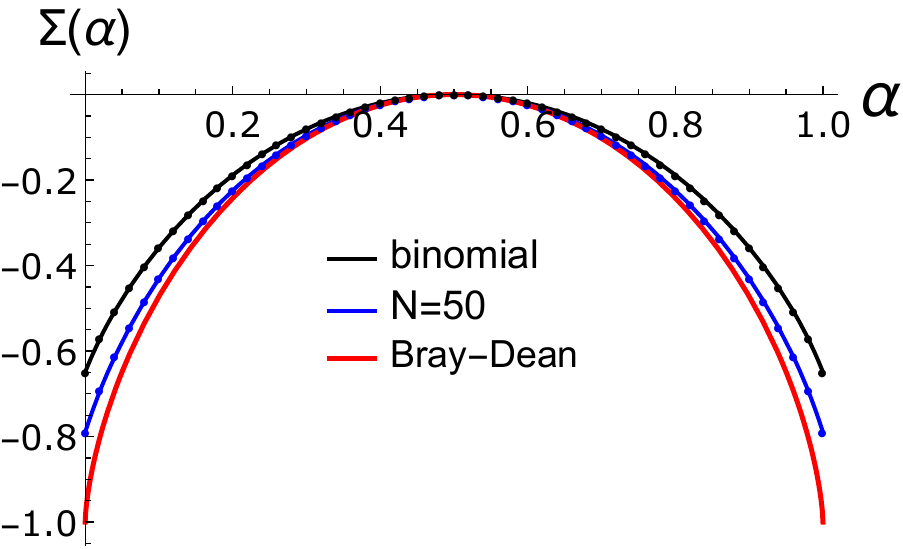} \\
   \includegraphics[width=4in]{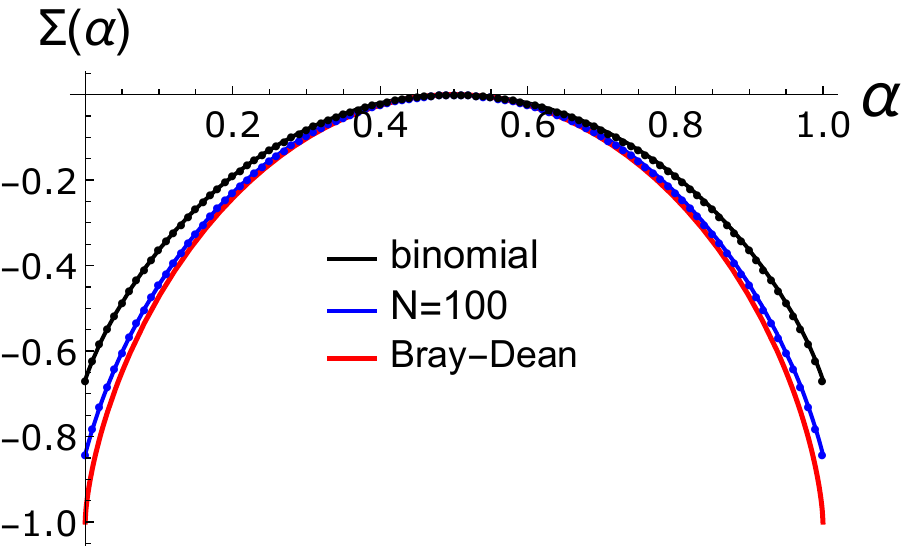} 
   \caption{Graphs of $\Sigma(\alpha)$ defined in \eqref{density} obtained from a binomial distribution in black (top curve), our exact calculation in blue (the middle curve) and the large-$N$ calculation of Bray-Dean in red (the bottom curve) for $N=10,  50$ and 100. We see that for small $N$ the binomial is a good approximation and while the $N=100$ case is close to the large $N$ limit \cite{Bray:2007tf} there is still a significant mismatch, when $\alpha$ is close to $0$ or $1$.}
   \label{fig:LargeNWide}
\end{figure}

The properties of stationary points of random Gaussian fields  in the large-$N$ limit have  been extensively studied, see for example Ref.~\cite{Bray:2007tf} and the references within. For landscape potentials $N$ is typically assumed to be ${\cal{O}}(100)$ and our next goal is to check the rate at which results for finite $N$  approach the large-$N$ limit. Adopting the notation Ref.~\cite{Bray:2007tf},  the Hessian of each stationary point of the potential can have between  $0$ to $N$ negative eigenvalues. Denote the number of stationary points whose Hessian has  $N \alpha$ negative eigenvalues by $\mathcal{N(\alpha)}$. By definition,  $\alpha$ is in the range $[0,1]$ and we express ${\cal N}(\alpha)$ in terms of ``complexity" $\Sigma(\alpha)$ via  
\bel{density}
	{\mathcal N}( \alpha) = e^{N \Sigma (\alpha)}~,
\ee
The complexity was calculated in \cite{Bray:2007tf} to be 
\bel{sigma}
	\Sigma(\alpha)= - \frac{\bar \lambda^2}{4 f''(0)}+ \text{normalization constant},
\ee
%
%
%
where $\bar \lambda$ is defined by
\bel{lambdabar}
	\frac2\pi\int_{\bar\lambda/2 \sqrt{f''(0)}}^1 dy \sqrt{1-y^2}= \alpha~. 
\ee 
Here $f$ is the correlation function defined in \eqref{Correlation}. We plotted the $\Sigma(\alpha)$ in Fig.\ref{fig:LargeNWide}. The center of $\Sigma(\alpha)$ fits well with a quadratic. We can calculate its width for $\alpha$ close to 1/2 (small $\bar\lambda$).
\be
	\bar \lambda^2= \pi^2 f''(0)\(\alpha-\frac12\)^2~,
\ee

\begin{figure}[htbp] 
   \centering
   \includegraphics[width=5in]{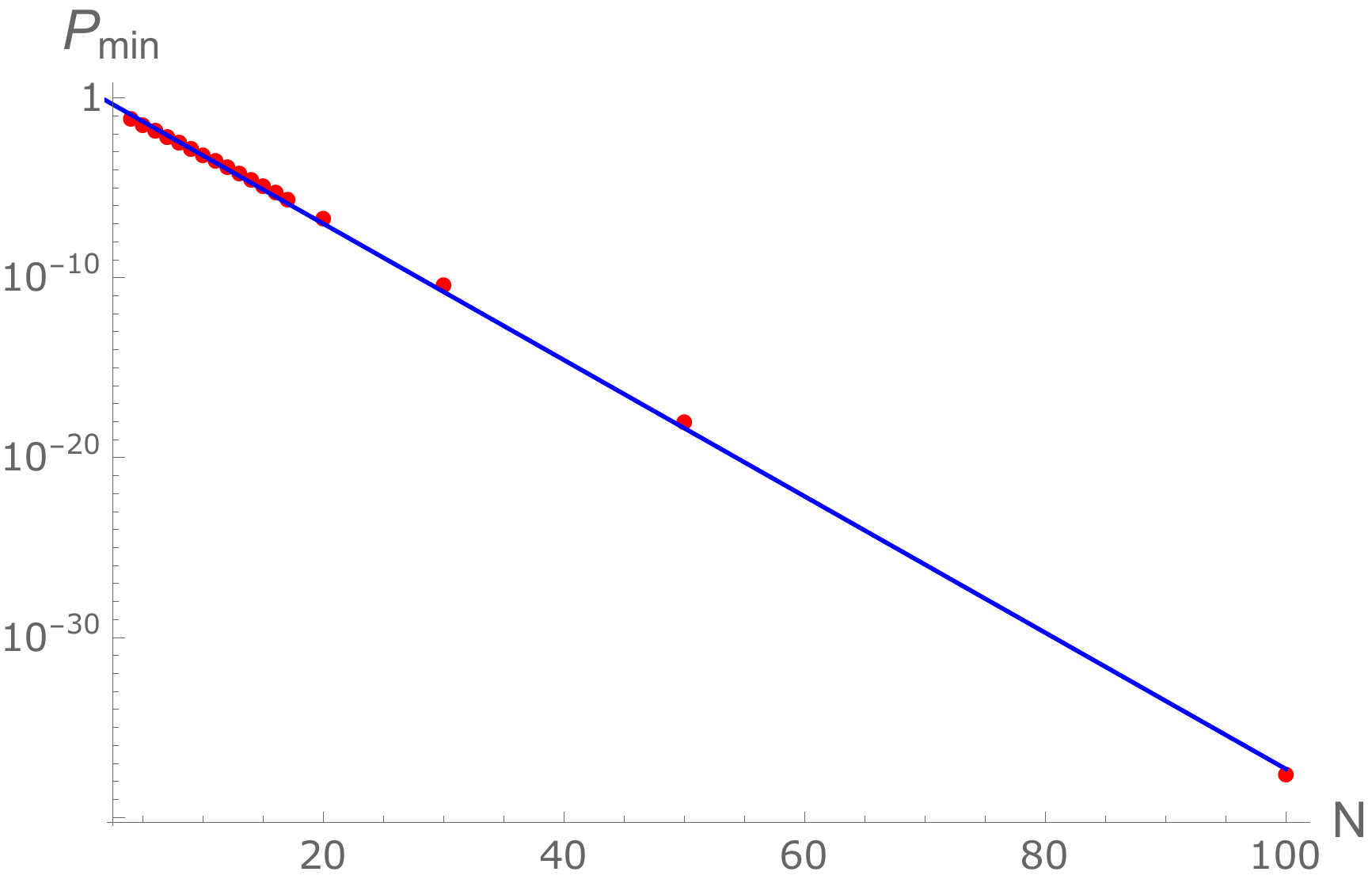} 
   \caption{Probability of finding a minimum among different stationary points as a function of $N$  the number of fields. It is very well approximated by $\ln P=1.32 - 0.87 N $ or $P \propto 2.39^{-N}$~.}
   \label{fig:FitMinNum}
\end{figure}

The complexity near the center is
\be
	\Sigma(\alpha)= -\frac{\pi^2}{4}\(\alpha-\frac12\)^2~.
\ee
 If one assumes a binomial distribution, this coefficient would be $2$ instead of $\pi^2/4$. This shows that the binomial distribution for large-$N$ values is not a good approximation. Our results at the center of the distribution are consistent with \cite{Bray:2007tf} but the discrepancy grows for small value of $\alpha$, corresponding to local minima. In Fig.\ref{fig:LargeNWide} we compare the complexity obtained from our exact results with the large-$N$ limit results of \cite{Bray:2007tf} and the binomial distribution for $N=10, 50$ and 100.  The binomial approximation  overestimates the density of minima relative to saddles, while the large-$N$ results of \cite{Bray:2007tf}  underestimates these values for finite $N$. However, in all cases the likelihood of a given extremum being a minimum decreases exponentially with $N$, rather than super-exponentially as would be the case if the Hessians were drawn from the Gaussian Orthogonal ensemble of random, symmetric matrices. This result is also consistent with the observations of Ref.~\cite{Masoumi:2016eqo}.  Finally, we plot the ratio of minima to stationary points computed from our evaluations of the \eqref{densityGeneral} in Fig.\ref{fig:FitMinNum} and it seems that it fits well by a line on a log scale.  

\section{Discussion and Conclusions}
\label{sec:conclusion}

We have taken steps towards quantifying expectations for the properties of landscape potentials embedded in theories of high energy physics, including string theory. Our overall approach is to begin with the {\em architecture\/} of the landscape, specifying  expectations for its global properties.  This paper  focuses on an apparently simple question; the relative numbers of minima and saddle points in generic landscapes. 
For $N=2$ we obtained strong results from Morse theory for general functions. At larger values of $N$ we begin from the default assumption that the landscape can be  modeled as a Gaussian random function and generalize methods used by Bardeen, Bond, Kaiser and Szalay to analyze primordial density fluctuations  \cite{Bardeen:1985tr} to treat this problem in $N$ dimensions.  

Our results demonstrate that for Gaussian random fields, saddles outnumber minima by a factor of roughly $2^N$. This is in contrast to analyses that treat the Hessians associated with extrema as random matrices, which  suggest that the ratio is closer to $\exp(-\alpha N^2)$ where $\alpha$ is a positive constant of order unity. The discrepancy arises because  Hessian matrices associated with a random function are not composed of independent and identically distributed elements. Consequently, while the present work  shares the fundamental philosophy of Ref.~\cite{Aazami:2005jf} that a sufficiently complex landscape can be modelled as a random distribution, the analysis must focus on the underlying {\em function\/}, and not the individual Hessians. Note also that for the Gaussian random functions the relative number of stationary points is independent of the power spectrum and depends only the ratios of integrals over the $\lambda_i$.

The analysis in this paper has established generic expectations for the relative numbers of different types of extrema in  Gaussian random function in many dimensions.  A  Gaussian random function with zero mean is arguably one of simplest possible specifications of the landscape architecture, and the distribution of potential cosmological constants $p(\Lambda)$ can be obtained in the large-$N$ limit \cite{Bray:2007tf}. We have shown that the total number of minima is only exponentially suppressed by increasing $N$, but $p(\Lambda|\Lambda \approx 0)$  will also depend  strongly on the values of  $\sigma_1$ and $\sigma_2$.\footnote{We may be able to extend our numerical methods to evaluate $p(\Lambda)$, but the integrand will contain $\sigma_1$ and $\sigma_2$ as well as $\Lambda$ so these computations will be less trivial than those performed here.}  We will save the full analysis of this scenario for future work. 

Beyond a  pure Gaussian random function, a more physically realistic landscape architecture might include both a random function and an overall potential arising from Planck-scale operators which is dominant at large values of $\bar\phi$. In this case, $p(\Lambda|\Lambda \approx 0)$ will depend on the overall position in the landscape. Conversely, the   layering   phenomenon described in Ref~\cite{CPA:CPA21422} implies that the minima  of $V$ will all be low-lying  for many possible landscape architectures in which case $p(\Lambda|\Lambda \approx 0)$ will be super-exponentially small  and the putative landscape cannot supply a single minimum in which $\Lambda$ is consistent with  observations or even our existence. Interestingly, in this situation the specific multiverse associated with the assumed form $\poten(\bar\field)$ would generate a strong prediction for $\Lambda$, and underlying hypothesis could be rejected with confidence. As a consequence, considerations of landscape architectures will -- at the very least  -- provide a sandbox for exploring  circumstances in which we can draw reliable inferences about multiverse scenarios, even in the presence of anthropic selection. From this starting point we can then consider landscape architectures which  are   physically reasonable while preserving, so far as possible, the overall brevity of the underlying specification.

\acknowledgements  

We are thankful to Thomas Bachlechner, William Handley, Lam Hui, Liam McAllister, Ken Olum and Alex Vilenkin  for valuable conversations. 
A.M is supported by a National Science Foundation grant 1518742. 

\appendix

\section{Random Gaussian ensemble of two fields}\label{sec:NEq2}

In this case the vector $\alpha$ has a simple form $\alpha_i=\{\gra_1, \gra_2, \hs_{11}, \hs_{22}, \hs_{12}\}$ and we have 
\be
 M=
\left(
\begin{array}{ccccc}
\frac{\sigma _1^2}{2} & 0 & 0 & 0 & 0 \\
 0 & \frac{\sigma _1^2}{2} & 0 & 0 & 0 \\
 0 & 0 & \frac{3 \sigma _2^2}{8} & \frac{\sigma _2^2}{8} & 0 \\
 0 & 0 & \frac{\sigma _2^2}{8} & \frac{3 \sigma _2^2}{8} & 0 \\
 0 & 0 & 0 & 0 & \frac{ \sigma _2^2}{8} \\
\end{array}
\right)~, 
\qquad K=\left(
\begin{array}{ccccc}
 \frac{2}{\sigma _1^2} & 0 & 0 & 0 & 0 \\
 0 & \frac{2}{\sigma _1^2} & 0 & 0 & 0 \\
 0 & 0 & \frac{3}{\sigma _2^2} & -\frac{1}{\sigma _2^2} & 0 \\
 0 & 0 & -\frac{1}{\sigma _2^2} & \frac{3}{\sigma _2^2} & 0 \\
 0 & 0 & 0 & 0 & \frac{8}{\sigma _2^2} \\
\end{array}
\right)~.
\ee
We can write the Hessian matrix in terms of two eigenvalues and a single Euler angle
\be
	\{ \zeta_{11}, \zeta_{12}, \zeta_{22}\}=\left\{\lambda _2 \sin ^2\theta+\lambda _1 \cos ^2\theta,(\lambda _1-\lambda_2) \sin \theta \cos \theta,\lambda _1 \sin ^2\theta+\lambda _2 \cos ^2\theta\right\}~.
\ee
Jacobian of this transformation is given by
\bel{det2Dim}
	J=\left| \frac{\partial\{ \hs_{11}, \hs_{12}, \hs_{22}\}}{\partial\{\lambda_1,\lambda_2, \theta\}} \right|= (\lambda_1-\lambda_2)~.
\ee
Because $\tilde Q$ in \eqref{alphaProb} is rotationally invariant, we evaluate it  at $\theta=0$.
\be
	\tilde Q=\frac{6 \left(\gra _1^2+\gra _2^2\right) \sigma_2^2+\left(9 \lambda _1^2-6 \lambda _2 \lambda _1+9 \lambda _2^2\right) \sigma_1^2}{6 \sigma_1^2 \sigma_2^2}.
\ee
The $\delta$ functions in \eqref{density1} sets $\gra_1=\gra_2=0$ which leads to
\be
	\tilde Q=\frac{3 \lambda _1^2-2 \lambda _2 \lambda _1+3 \lambda _2^2}{2\sigma_2^2}~.
\ee
The probability density of $\alpha_i$ simplifies immensely 
\be
	p(\alpha_i)=\frac{1}{(2\pi)^{5/2}}\frac{16}{\sigma ^2_1 \sigma ^3_2\sqrt{3}} \exp \(-\frac{3 \lambda _1^2-2 \lambda _2 \lambda _1+3 \lambda _2^2}{2\sigma_2^2} \)~.
\ee
Now we can calculate the number of minima, maxima and saddle points (keeping in mind $\lambda_1\ge \lambda_2$ and $\det (\hs_{ij})=\lambda_1 \lambda_2$ and a factor of $\frac12$ for the double counting in rotation group).
\bea
	n_i&=& \frac12 \int_0^{2\pi} d\theta \int_{-\infty}^\infty d\lambda_1\int_{-\infty}^{\lambda_1} d\lambda_2  \lambda_1 \lambda_2 (\lambda_1-\lambda_2)\frac{1}{(2\pi)^{5/2}}\frac{16}{\sigma ^2_1 \sigma ^3_2\sqrt{3}} \exp \(-\frac{3 \lambda _1^2-2 \lambda _2 \lambda _1+3 \lambda _2^2}{2\sigma _2^2} \) \theta_H \nn
	&=&\frac{1}{(2\pi)^{3/2}}\frac{8\sigma ^2_2}{\sigma ^2_1 \sqrt{3}}\int_{-\infty}^\infty d\lambda_1\int_{-\infty}^{\lambda_1} d\lambda_2  \lambda_1 \lambda_2 (\lambda_1-\lambda_2) \exp \(-\frac{3 \lambda _1^2-2 \lambda _2 \lambda _1+3 \lambda _2^2}{2} \) \theta_H
\eea
where $\theta_H$ is the appropriate $\theta$ function. It is easy to check that this integral vanishes without a $\theta$ function as expected form the Morse theory. We get 
\bea
	\{n_{\rm min}, n_{\rm saddle}, n_{\rm max}\}&=&\frac{\sigma_2^2}{24 \pi \sigma_1^2}\{1, 2, 1 \}~,
\eea
This is coincidently a binomial distribution for the relative number of maxima, minima and saddle points. The relative numbers from a random Gaussian ensemble is 
\be
	\{n_{\rm min}, n_{\rm saddle}, n_{\rm max}\}\propto\{1, 2(1+\sqrt2), 1 \}~.
\ee

\section{N=3}\label{sec:Neq3}
In this case we define the following vector 
\bel{alpha3}
	\alpha=\left\{\eta _1,\eta _2,\eta _3,\zeta _{11},\zeta _{22},\zeta _{33},\zeta _{23},\zeta _{13},\zeta _{12}\right\}~.
\ee
In this basis the matrices $M$ and $K$ are given by
\be
	M=\left(
\begin{array}{ccccccccc}
 \frac{\sigma _1^2}{3} & 0 & 0 & 0 & 0 & 0 & 0 & 0 & 0 \\
 0 & \frac{\sigma _1^2}{3} & 0 & 0 & 0 & 0 & 0 & 0 & 0 \\
 0 & 0 & \frac{\sigma _1^2}{3} & 0 & 0 & 0 & 0 & 0 & 0 \\
 0 & 0 & 0 & \frac{\sigma _2^2}{5} & \frac{\sigma _2^2}{15} & \frac{\sigma _2^2}{15} & 0 & 0 & 0 \\
 0 & 0 & 0 & \frac{\sigma _2^2}{15} & \frac{\sigma _2^2}{5} & \frac{\sigma _2^2}{15} & 0 & 0 & 0 \\
 0 & 0 & 0 & \frac{\sigma _2^2}{15} & \frac{\sigma _2^2}{15} & \frac{\sigma _2^2}{5} & 0 & 0 & 0 \\
 0 & 0 & 0 & 0 & 0 & 0 & \frac{\sigma _2^2}{15} & 0 & 0 \\
 0 & 0 & 0 & 0 & 0 & 0 & 0 & \frac{\sigma _2^2}{15} & 0 \\
 0 & 0 & 0 & 0 & 0 & 0 & 0 & 0 & \frac{\sigma _2^2}{15} \\
\end{array}
\right),
K=\left(
\begin{array}{ccccccccc}
 \frac{3}{\sigma _1^2} & 0 & 0 & 0 & 0 & 0 & 0 & 0 & 0 \\
 0 & \frac{3}{\sigma _1^2} & 0 & 0 & 0 & 0 & 0 & 0 & 0 \\
 0 & 0 & \frac{3}{\sigma _1^2} & 0 & 0 & 0 & 0 & 0 & 0 \\
 0 & 0 & 0 & \frac{6}{\sigma _2^2} & -\frac{3}{2 \sigma _2^2} & -\frac{3}{2 \sigma _2^2} & 0 & 0 & 0 \\
 0 & 0 & 0 & -\frac{3}{2 \sigma _2^2} & \frac{6}{\sigma _2^2} & -\frac{3}{2 \sigma _2^2} & 0 & 0 & 0 \\
 0 & 0 & 0 & -\frac{3}{2 \sigma _2^2} & -\frac{3}{2 \sigma _2^2} & \frac{6}{\sigma _2^2} & 0 & 0 & 0 \\
 0 & 0 & 0 & 0 & 0 & 0 & \frac{15}{\sigma _2^2} & 0 & 0 \\
 0 & 0 & 0 & 0 & 0 & 0 & 0 & \frac{15}{\sigma _2^2} & 0 \\
 0 & 0 & 0 & 0 & 0 & 0 & 0 & 0 & \frac{15}{\sigma _2^2} \\
\end{array}
\right)~.
\ee
The most general rotation in three dimension is given in terms of the Euler angles $\xi, \beta$ and $\gamma$, where $0\le \xi,\gamma < 2\pi$ and $0\le \beta <\pi$~. 
The rotation is given by
\be
	R_{\rm Euler}=\left(
\begin{array}{ccc}
 \cos \xi\cos\beta-\cos\beta \sin \xi\sin\gamma & \cos\gamma \sin \xi+\cos \xi\cos\beta \sin\gamma & \sin\beta \sin\gamma \\
 -\cos\beta \cos\gamma \sin\xi-\cos \xi\sin\gamma & \cos \xi\cos\beta \cos\gamma-\sin \xi\sin\gamma & \cos\gamma \sin\beta \\
 \sin \xi\sin\beta & -\cos \xi\sin\beta & \cos\beta \\
\end{array}
\right)
\ee
Again we rewrite the matrix of second derivatives in terms of the eigenvalues and rotation angles 
\be
\left(
\begin{array}{ccc}
 \xi _{11} & \xi _{12} & \xi _{13} \\
 \xi _{12} & \xi _{22} & \xi _{23} \\
 \xi _{13} & \xi _{23} & \xi _{33} \\
\end{array}
\right)=  R_{\rm Euler} \left(
\begin{array}{ccc}
 \lambda _1 & 0 & 0 \\
 0 & \lambda _2 & 0 \\
 0 & 0 & \lambda _3 \\
\end{array}
\right)R^T_{\rm Euler}~.
\ee
We can easily calculate the Jacobian using Mathematica to get 
\be
	J=\left| \frac{\partial\{ \zeta_{11}, \zeta_{22}, \zeta_{33}, \zeta_{23}, \zeta_{13}, \zeta_{23}\}}{\partial\{\lambda_1,\lambda_2, \lambda_3, \xi, \beta, \gamma\}} \right|= (\lambda_1-\lambda_2)(\lambda_1-\lambda_3) (\lambda_2-\lambda_3) \sin \beta~.
\ee
To get the distribution of $\alpha_i$ vectors, we again calculate  $\tilde Q$ 
\be 
	\tilde Q= \frac{3 (2 \zeta _{11}^2-(\zeta _{22}+\zeta _{33}) \zeta _{11}+2 \zeta _{22}^2+2 \zeta _{33}^2+5 (\zeta _{12}^2+\zeta _{13}^2+\zeta _{23}^2)-\zeta _{22} \zeta _{33})}{2\sigma _2^2}+\frac{3 (\eta _1^2+\eta _2^2+\eta _3^2)}{2\sigma _1^2}
\ee
We evaluate this at $\eta_1=\eta_2=\eta_3=0$ because of the delta functions and we use the spherical symmetry to set  $\xi=\beta=\gamma=0$ which makes 
$\zeta_{12}=\zeta_{13}=\zeta_{23}=0$ and $\zeta_{11}=\lambda_1, \zeta_{22}=\lambda_2$ and $\zeta_{33}=\lambda_3$.
This gives 
\be
	\tilde Q=\frac{3}{4\sigma_2^2} \(5 \sum\lambda_i^2 -(\sum \lambda_i)^2 \)~.
\ee
From here the density of different type of minima is given by 
\bea
	N_i&=& \frac{5^{5/2}3^{9/2} }{2 \sigma_1^3\sigma_2^6 (2\pi)^{9/2}}\int_0^{2\pi}d\alpha \int_0^{2\pi}d\gamma  \int_0^\pi d\beta \sin \beta\int d\eta_1 d\eta_2 d\eta_3 \delta(\eta_1)\delta(\eta_2)\delta(\eta_3)\nn
	&&\int d \lambda_1 d\lambda_2 d\lambda_3 e^{-\tilde Q} (\lambda_1-\lambda_2)(\lambda_1-\lambda_3) (\lambda_2-\lambda_3) \lambda_1 \lambda_2 \lambda_3 \theta_H \nn
	&=&9 \(\frac{15}{2\pi} \)^{5/2}\frac{\sigma_2^3}{\sigma_1^3 }\int \Pi (\lambda_i d \lambda_i)   (\lambda_1-\lambda_2)(\lambda_1-\lambda_3) (\lambda_2-\lambda_3) \exp\[-\frac{3}{4} \(5 \sum\lambda_i^2 -(\sum \lambda_i)^2 \)\] \theta_H~. \nn
\eea
Again $\theta_H$ determines the type of stationary point we want to calculate. Again we can show that without the $\theta$ function this integral vanishes as expected from Morse theory. Now we evaluate the density of different type of stationary points. 
\bea
	n_0&=&n_3=\frac{29 \sqrt{15}-18 \sqrt{10}}{450 \pi ^2} \frac{\sigma_2^3}{\sigma_1^3}~, \nn
	n_1&=&n_2=\frac{29 \sqrt{15}+18 \sqrt{10}}{450 \pi ^2} \frac{\sigma_2^3}{\sigma_1^3}~.
\eea
The ratio $n_1/n_0$ is approximately 3.05 which is close to but still distinct from the value of 3 that we would expect from binomial distribution.  

\section{Distribution of different types of stationary points for $N=50$}\label{sec:NEq50}
In this section we present the data for the chance of finding a stationary point with $n$ negative eigenvalues from a set of stationary points. Because we expect $P(i)=P(50-i)$ from the symmetry $U \rightarrow -U$ we only show the data for $n$ up to 25 in Table \ref{table:Pn}. 
\begin{table}
\begin{tabular}{|c|c|c|c|c|c|c|c|c|c|c|}
\hline
$n$ &  0& 1& 2& 3& 4& 5 & 6   \\ \hline
$P(n)$ &  $8.43\times 10^{-19}$&$1.07\times 10^{-16}$&$5.69\times 10^{-15}$&$1.77\times 10^{-13}$&$3.65\times 10^{-12}$&$5.59\times 10^{-11}$&$6.51\times 10^{-10}$ \\ \hline
$n$  &7&8 &9&10&11&12&13 \\ \hline
$P(n)$&$ 6.11\times 10^{-9}$&$ 4.67\times 10^{-8} $ & $2.98\times 10^{-7}$&$1.64\times 10^{-6}$&$7.73\times 10^{-6}$&$3.18\times 10^{-5}$&$1.14\times 10^{-4}$\\ \hline 
$n$&14&15&16& 17& 18 &19 & 20   \\ \hline
$P(n)$&$3.64\times 10^{-4}$&$1.04\times 10^{-3}$&$2.62\times 10^{-3}$&$5.95\times 10^{-3}  $& $1.22 \times 10^{-2}$ & $2.28\times 10^{-2}$ &$ 3.83\times 10^{-2}$ \\ \hline
$n$ & 21 & 22 &23 & 24 & 25&26 & 27\\ \hline
 $P(n)$ & $5.81\times 10^{-2}$& $8.00\times 10^{-2}$& $0.101$& $0.117$& $0.122$&$0.116$&0.101 \\ \hline
\end{tabular}
\caption{The values of $P(n)$ defined above for $N=50$. From here we can reconstruct $\Sigma(\alpha)$ shown in Fig.\ref{fig:LargeNWide}. These numbers are obtained from a Monte-Carlo integration which is accurate to 1 percent.}
 \label{table:Pn}
\end{table}

\bibliography{random}
\end{document}